# Spin-orbit interactions in noncentrosymmetric plasmonic crystals probed by site-selective cathodoluminescence spectroscopy

Masoud Taleb[1], Mohsen Samadi[1], Fatemeh Davoodi[1], Maximilian Black[1], Janek Buhl[2], Hannes Lüder[2], Martina Gerken[2], and Nahid Talebi[1,3*]

[1]Institute of Experimental and Applied Physics, Kiel University, 24098 Kiel, Germany

[2]Integrated Systems and Photonics, Faculty of Engineering, Kiel University, 24143 Kiel, Germany

[3]Kiel, Nano, Surface, and Interface Science, Kiel University, 24098 Kiel, Germany

[*]E-Mail: talebi@physik.uni-kiel.de

## Abstract

The study of spin-orbit coupling (SOC) of light is crucial to explore the light-matter interactions in sub-wavelength nanostructures with broken symmetries. In noncentrosymmetric photonic crystals, the SOC results in the splitting of the otherwise degenerate energy bands. Herein, we explore the SOC in a noncentrosymmetric plasmonic crystal, both theoretically and experimentally. Cathodoluminescence (CL) spectroscopy combined with the numerically calculated photonic band structure reveals an energy band splitting that is ascribed to the broken symmetries in the noncentrosymmetric plasmonic crystal. By shifting the impact position of the electron beam throughout a unit cell of the plasmonic crystal, we show that the emergence of the energy band splitting strongly depends on the excitation position of the surface plasmon (SP) waves on the crystal. Moreover, we exploit angle-resolved CL and dark-field polarimetry to demonstrate polarization-dependent scattering of SP waves interacting with the plasmonic crystal. The scattering direction of a given polarization is determined by the transverse spin angular momentum inherently carried by the SP wave, which is in turn locked to the direction of SP propagation. Our study gives insight into the design of novel plasmonic devices with polarization-dependent directionality of the Bloch plasmons. We expect spin-orbit plasmonics will find much more scientific interests and potential applications with the continuous development of nanofabrication methodologies and uncovering new aspects of spin-orbit interactions.

# Introduction

Spin-orbit interactions have raised a great deal of interest among the condensed-matter physics and material engineering communities owing to their potential applications for the design of the topological states of matter[1,2] in the developing field of spintronics[3]. The main goal for this area of research is to generate and control spin-polarized states by manipulation of inversion and time-reversal (TR) symmetries in solids. The Rashba[4,5] and Dresselhaus[6] effects serve as examples of the SOC phenomena occurring when the inversion symmetry of the crystal is broken. Even in the absence of an external magnetic field, these SOC effects induce an effective momentum ($\hbar\mathbf{k}$) – dependent magnetic field $\mathbf{\Omega}(\mathbf{k})$ and the SOC dispersion can be described by the Hamiltonian:

$$H_{\text{SOC}} = \mathbf{\Omega}(\mathbf{k}) \cdot \boldsymbol{\sigma} \tag{1}$$

where $\sigma$ denotes the vector of the Pauli spin matrices and $\mathbf{\Omega}(\mathbf{k})$ depends on the spatial symmetry of the system[7]. The SOC Hamiltonian splits the spin-degenerate energy bands by shifting the band structures of the electrons with opposite spins to opposite directions in momentum space (Figure 1a). In addition to the bulk Rashba and Dresselhaus effects, another type of SOC has been observed in nonmagnetic crystals with local inversion asymmetry[8,9]. It has been shown that the SOC is highly localized to atomic sites, and therefore the local inversion asymmetry in one sector of a crystal gives rise to a local spin polarization. The total spin polarization of the crystal is then determined by the superposition of the local spin polarizations induced at individual sectors. Interestingly, the interaction between distinct sectors leads to an additional band splitting and emergence of non-degenerate energy bands with a mixture of spin polarizations localized in each of the sectors.

The optical counterpart of the Rashba-type SOC was presented in a series of studies[10–13] wherein right (left)-handed circular polarization of light plays the role of the spin-up (down) state. It was shown that breaking the inversion symmetry in plasmonic crystals and metasurfaces triggers the spin-orbit interactions and results in spin-split dispersion similar to the electronic Rashba effect. Additionally, quantum spin Hall effect (QSHE) of light implies that guided or surface waves with opposite transverse spin angular momenta (SAM) propagate in opposite in-plane directions[14–17]. Upon interacting with an arbitrary scatterer, a surface wave of a given circular polarization is scattered along a certain direction determined by the state of polarization and the direction of the surface wave impinging at the scatterer. The QSHE provides a unique method for SAM-to-direction coupling and unidirectional excitation of guided or surface modes in plasmonic crystals and metasurfaces.

The photonic Rashba effect displaces the otherwise degenerate spin-up and spin-down bands along the momentum axis. There exists a second SOC effect, i.e. the Zeeman effect, which moves the bands along the energy axis. While the conventional Zeeman effect is necessarily induced by applying an external magnetic field[18], non-magnetic Zeeman-type splitting is observed in 2D transition-metal dichalcogenides (TMDCs)[19,20] and 3D noncentrosymmetric materials with specific symmetry point groups[9,21]. The photonic counterpart of the Zeeman-type effect though has not been reported yet. In this paper, we present a plasmonic crystal with broken inversion symmetry that exhibits SOC as well as split energy bands owing to simultaneous existence of the Rashba-type and Zeeman-type effects (Figure 1a). The photonic crystal structure proposed here is composed of a multilayer SiO$_2$/ITO/Au configuration with a unit cell that exhibits varying ITO layer heights (Figure 1b). We demonstrate spin-dependent direction splitting of the surface plasmon modes caused by the QSHE of light, using the angle-resolved dark-field and cathodoluminescence polarimetry. For the CL polarimetry measurements (Figure 1b), we excite the sample with subwavelength control of the electron impact position and analyze the polarization state of the angle-resolved CL patterns (see Materials and Methods for the details). With a subwavelength shift of the electron impact position from a position in which the plasmonic crystal is viewed centrosymmetric to another point where the structure lacks inversion symmetry, a splitting is observed in the angle-resolved patterns. The site-selective emergence of the band splitting provides a unique feature of the angle-resolved CL polarimetry that cannot be achieved via dark-field measurements. The energy-momentum dispersion of the plasmonic crystal measured by cathodoluminescence spectroscopy and the numerically calculated band diagram both confirm the emergence of SOC-induced band-splitting in our presented noncentrosymmetric plasmonic crystal. The angular momentum vorticities calculated for each of the split modes show rotating behaviors in opposite directions due to the spin-orbit interactions in this structure.

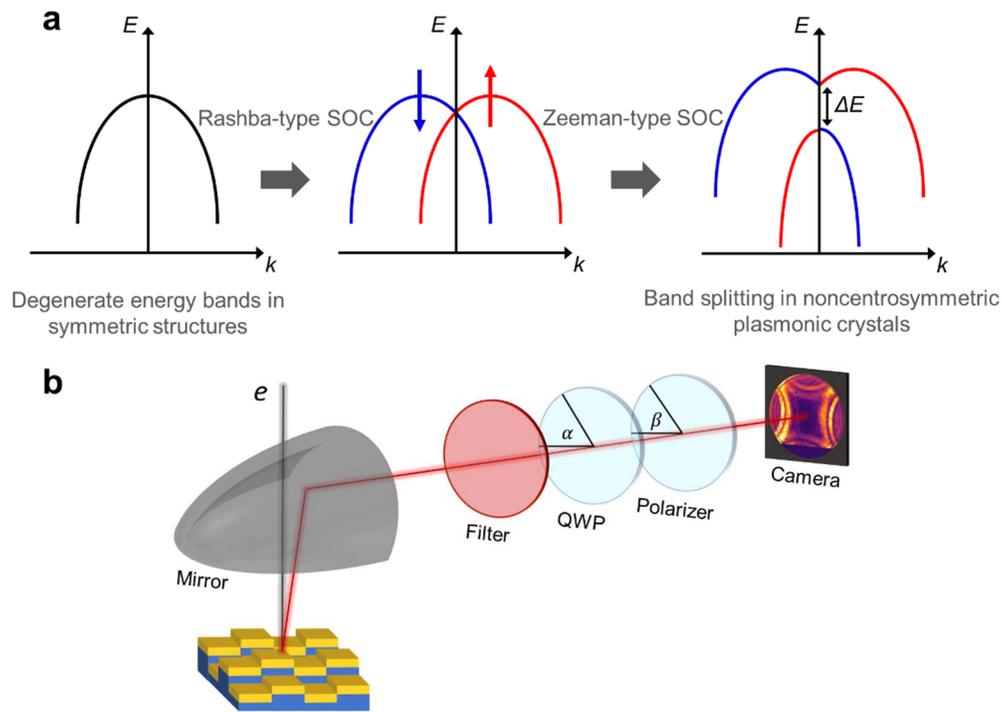

**Figure 1.** Spin-Orbit Coupling (SOC) and energy band splitting in plasmonic crystal. (a) Rashba-type and Zeeman-type spin-orbit coupling occur in noncentrosymmetric plasmonic crystal owing to inversion symmetry breaking. (b) Experimental setup for angle-resolved CL polarimetry equipped with a filter, a quarter wave plate (QWP) and a linear polarizer to measure the polarization state of the CL emission of the sample. The sample is excited by an electron beam and the resulting CL emission is directed to a camera by a parabolic mirror (Projected on the CCD camera is the $S_0$ Stokes parameter of the lattice at $\lambda_0$ = 800 nm ($E$ = 1.55 eV)).

## Results

In order to investigate the behavior of the noncentrosymmetric plasmonic crystal systematically, we provide a comparison between the following cases: (i) 1D ITO grating patterned on a glass substrate (Figure 2a), (ii) Au-coated 1D ITO grating patterned on a glass substrate (Figure 2d), and (iii) 2D noncentrosymmetric plasmonic crystal (Figure 2g). In case (i), a 140-nm thick ITO layer is patterned to form a 1D grating with the periodicity of $p$ = 360 nm and the step height of $h_{step}$= 45 nm. To create the structure

(ii), a thin film of gold ($h_{Au}$ = 40 nm) is deposited on the realized 1D ITO grating. The latter structure, i.e. the 2D noncentrosymmetric plasmonic crystal, is created by two consecutive etching steps with different depths in perpendicular directions followed by the deposition of a 40-nm thick Au layer on the patterned ITO (see the Supporting Information for the details of the fabrication method)[22]. The periodicity of the noncentrosymmetric plasmonic crystal is $p$ = 500 nm. As illustrated in the inset of Figure 2g, each unit cell of the realized noncentrosymmetric plasmonic crystal consists of four sectors with different ITO heights ($h_1$ = 65 nm, $h_2$ = 90 nm, $h_3$ = 115 nm, $h_4$ = 140 nm) coated with gold. We numerically calculated the spatial distribution of the z-component of the electric field ($E_z$) and measured the Stokes parameters with Dark-field polarimetry to analyze the polarization state of the light that is scattered from the aforementioned structures.

## Electric field calculation

The z-component of the electric field ($E_z$) within a single unit cell of the structures (i)-(iii) are shown in Figure 2b, 2e, 2h, overlaid with arrows indicating the electric field polarization at each position. In samples (i) and (ii), since the structure is homogenous in y-direction, the electric field is calculated in x-z plane to reduce the computation time. The 1D grating couples the incident light to a partially evanescent and partially propagating wave, i.e. the waveguide mode of the ITO layer (in case (i)) and the surface plasmon mode at the interface of Au and ITO (in case (ii)), both propagating along the spatial modulation of the grating, i.e. ±x-direction, and with an evanescent tail along the z-direction. For a wave that propagates along +x-direction and decays exponentially along +z-direction, one can generally write the electric field as follows (see the Supporting Information for the field calculation details)[16,23]:

$$\mathbf{E} = \frac{A}{\sqrt{1+|m|^2}} \left( -im \frac{\kappa}{k_x} \mathbf{\bar{x}} + \frac{k}{k_x} \mathbf{\bar{y}} + \mathbf{\bar{z}} \right) \exp(i k_x x - \kappa z) \qquad (2)$$

Here $\mathbf{k} = k_x \mathbf{\bar{x}} + i\kappa \mathbf{\bar{z}}$ is the complex wavevector, whereas $k_x$ and $\kappa = \sqrt{k_x^2 - k^2}$ are the longitudinal wave number along x-direction and the exponential decay rate along z-direction, respectively. $\mathbf{\bar{\alpha}}$ with $\alpha \in$ (x, y, z) is the unit vector along the specified direction. The complex number $m$, as defined in the Supporting Information, describes the polarization state of the wave. In contrast to propagating waves in free space with pure transverse polarizations, the wave described by equation (2) possesses an imaginary longitudinal electric field component that induces rotation of the electric field in x-z plane resulting in an extraordinary transverse spin[23,24] that can be written as[16]:

$$\mathbf{S}_\perp = \frac{\text{Re } \mathbf{k} \times \text{Im } \mathbf{k}}{(\text{Re } \mathbf{k})^2} \qquad (3)$$

As opposed to the longitudinal SAM, the transverse SAM is orthogonal to the wavevector of the propagating wave and, interestingly, the direction of the transverse SAM is locked to the propagation direction of the wave[16,25–27]. As can be observed in Figure 2b and 2e, the electric field rotates within the x-z plane, thereby generating a transverse SAM along the orthogonal y-direction. The spin-momentum locking phenomenon is more visible in the field profile of the 3D plasmonic crystal presented in Figure 2h, where oppositely propagating waves (+x and −x) carry transverse SAMs with opposite directions (+y and −y) and orthogonally propagating waves (+x and +y) possess SAMs in orthogonal directions (+y and −x). This property of the propagating waves with evanescent tails will become more obvious by measuring the polarization states of the scattered light from the structures, as it is described below.

## Dark-field polarimetry

When an evanescent wave propagates in a certain direction (e.g. +x), it is scattered to the free space upon interaction with the periodic structure. Owing to the non-zero transverse SAM of the evanescent wave, the scattered light is decomposed into opposite circular polarizations (RCP and LCP) emitting in opposite directions (−y and +y). An evanescent wave propagating in the −x direction, on the other hand, carries a transverse SAM in the opposite direction. Thus, directional splitting of the RCP and LCP components of the scattered light can be observed in the opposite direction (+y and −y). To gain a better understanding of the spin-momentum locking phenomena, we analyzed the polarization state of the light scattered from each of the previously mentioned structures via dark-field polarimetry. For this purpose, we placed a quarter wave plate (QWP) followed by a linear polarizer at the detection path of a dark-field microscopy setup and measured the Stokes parameters for each sample (see Materials and Methods for more details about dark-field polarimetry).

The angular patterns of $S_1$ and $S_3$ Stokes parameters corresponding to the different samples are presented in Figure 2c, 2f, 2i. The Stokes parameters obtained from the 1D gratings (Figure 2c and 2f) demonstrate that light is coupled to guided transverse electric (TE) modes propagating along ±x-direction with clear splitting of the RCP and LCP components (indicated by red and blue colors in $S_3$, respectively). Comparison of the Stokes parameters obtained from the 1D ITO grating without/with Au (Figure 2c and 2f) reveals that guided modes in both samples exhibit a similar spin-dependent direction splitting behavior. Nevertheless, excitation of surface plasmon waves in the latter case (Figure 2f) enhances the intensity of the scattered light and emphasizes the splitting.

In the 2D plasmonic crystal, light is mostly coupled to transverse magnetic (TM) surface plasmon waves that propagate along orthogonal in-plane directions, i.e. ±x and ±y. However, an almost directional propagation of the surface plasmon waves is observed in the corresponding angle-resolved Stokes parameters (Figure 2i), where the modes propagating along x-direction are more visible than those propagating in y-direction. To better understand the origin of this nearly directional propagation, one has to notice the fabrication process of the sample (explained in the Supporting Information), wherein an ITO layer was first etched to form a 1D grating with spatial modulation along x-direction and a groove height of 25 nm. Then, a second round of etching was carried out along y-direction with a different etching depth of 50 nm. As a result, the groove height of the realized 2D ITO grating along y-direction is not of the same value as that along x-direction ($h_g^x$ = 25 nm, $h_g^y$ = 50 nm). Deposition of a 40-nm thick Au film on the resulting 2D ITO grating produces a 2D plasmonic structure in which the metallic layer is continuous along x-direction, whereas formation of discontinuities along y-direction hinders the propagation of surface plasmon waves in the latter direction. The angle-resolved pattern of the $S_3$ Stokes parameter (Figure 2i) shows spin-dependent direction splitting of the modes propagating either along x or y direction. The measurements here were based on the RGB camera installed in our setup for Fourier imaging, thus accessing to the behavior of the structures in the longer wavelength ranges were not possible. Hence, we use CL spectroscopy to gain more insight into the broadband and site-specific behaviors of the noncentrosymmetric structure.

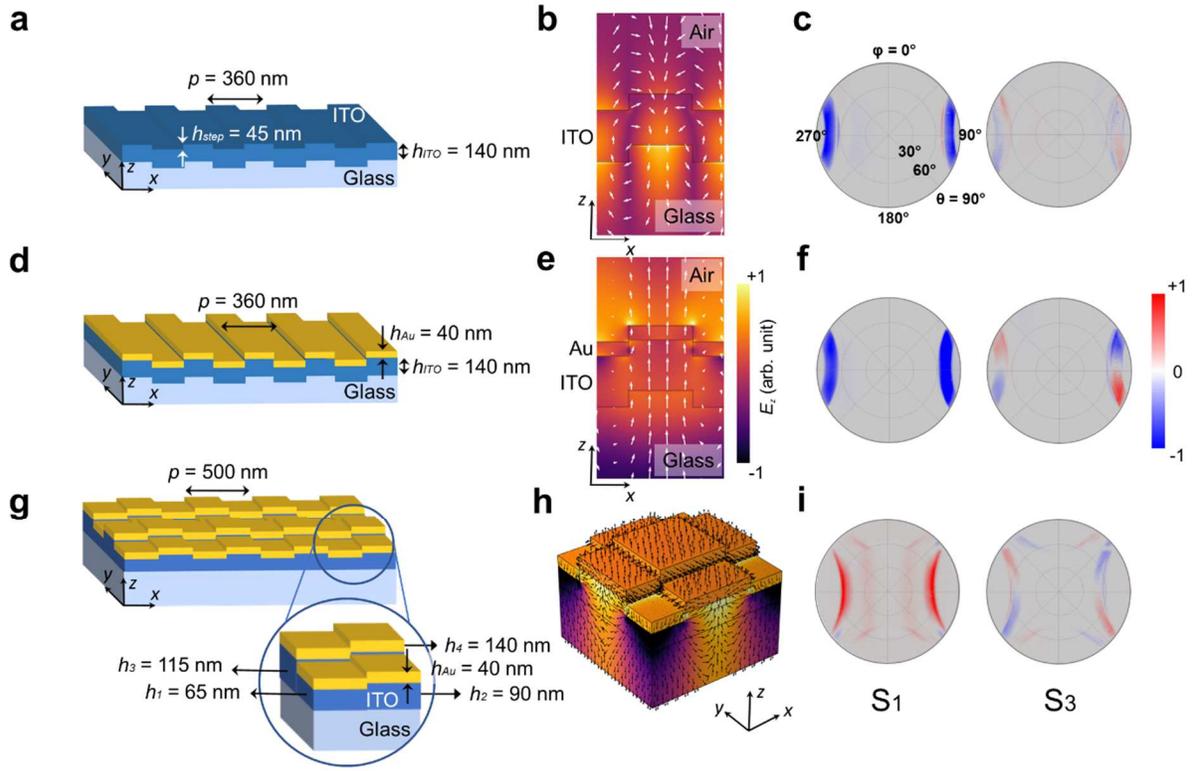

**Figure 2.** Spin-Momentum locking of evanescent waves in photonic and plasmonic crystals. (a, d, g) Schematic representation of (a) 1D ITO grating patterned on a glass substrate, (d) Au-coated 1D ITO grating patterned on a glass substrate, and (g) 2D noncentrosymmetric plasmonic crystal. The inset in (g) indicates the zoomed in view of a unit cell of the 2D noncentrosymmetric plasmonic crystal. (b, e, h) Spatial distribution of the z-component of the electric field ($E_z$) within a single unit cell of the nanostructures shown in panels (a, d, g) calculated at $\lambda_0$ = 550 nm ($E \approx$ 2.25 eV) in (b, e) and $\lambda_0$ = 600 nm ($E \approx$ 2.07 eV) in (h). The arrows show the electric field polarization at each position. (c, f, i) Stokes parameters $S_1$ and $S_3$ corresponding to the structures shown in panels (a, d, g), measured at $\lambda_0$ = 550 nm ($E \approx$ 2.25 eV) in (c, f) and $\lambda_0$ = 600 nm ($E \approx$ 2.07 eV) in i.

## Cathodoluminescence spectroscopy and polarimetry

To gain more insight into the optical behavior of the proposed noncentrosymmetric plasmonic crystal, we have measured the energy-momentum dispersion of the CL emission from the sample along the direction indicated by a red arrow in the inset of Figure 3a. The spot size of the electron beam is fixed to cover the whole unit cell of the structure. The dispersion diagram is acquired by placing a 1D slit in the optical path of the CL emission to select a specific azimuthal angle and a range of polar angles from 0 to 90 degree. The filtered CL light is dispersed by means of a grating onto the screen of a CCD camera[28]. The energy-momentum dispersion map shows a clear energy band splitting within a broad range of momentum and energy values (E < 1.8 eV, λ > 689 nm). Note that the dark zone in the low-k region of the energy-momentum CL map in Figure 3a is due to the hole inserted in the hyperbolic mirror, to allow the electron beam to pass through the mirror. We also plotted the CL spectrum at a fixed in-plane wavevector that is indicated by a vertical dashed line on the energy-momentum map and marked the spectral positions of the two peaks by colored arrows (Figure 3b). This double-peak phenomenon, wherein two parallel bands are observed over a wide energy and momentum range, only occurs in the 2D noncentrosymmetric plasmonic crystal and the CL response of a symmetric structure, e.g. a 1D grating or a 2D centrosymmetric plasmonic crystal) does not demonstrate an energy-splitting behavior. Thus, we attribute this phenomenon to the inversion asymmetry of the structure. In the followings, we will show that the differences in these two optical modes are due to the differences in the angular momenta of light.

To investigate the polarization state of the CL emission from the sample, we determine the Stokes parameters $S_0$, $S_1$ and $S_3$ by using the angle-resolved CL polarimetry setup demonstrated in Figure 1b. The Stokes parameters were acquired at two different electron impact positions indicated schematically in Figure 3c and 3e: (1) where the ITO thickness is $h_1$= 65 nm (Figure 3c) and (2) at the crossing point of the four sectors of a unit cell (Figure 3e). The resulting CL emission is filtered at a certain energy $E$ = 1.55 eV ($\lambda$ ≈ 800 nm). When the electron beam traverses the sample at position 1 (Figure 3c), the measured Stokes parameters exhibit spin-dependent direction splitting of the surface plasmon modes propagating along $x$ or $y$ direction (Figure 3d), which is in total agreement with the results achieved by the dark-field polarimetry method. Considering the mode propagating in the +$x$-direction, demonstrated by the semicircle at the right side of the angular $S_3$ pattern, the state of circular polarization (spin) flips sign by altering the azimuthal angle $\varphi$ at a certain polar angle, e.g. $\theta$ = 60°. We ascribe this azimuthal spin-splitting behavior to the spin-momentum locking phenomenon in plasmonic waves. As mentioned before, this phenomenon is a universal property of the plasmonic waves and appears regardless of the symmetry condition of the system. Excitation of the sample at the crossing point of the four sectors of a unit cell (Figure 3e), on the other hand, leads to the emergence of two non-degenerate modes as previously demonstrated in Figure 1a. It can be observed clearly in the Stokes parameters $S_0$, $S_1$ and $S_3$ angle-resolved patterns (Figure 3f) that an additional splitting of the modes propagating in the $x$ and $y$ directions appears. A closer look at the $S_3$ parameter provides further information about the state of circular polarization (spin) of the split modes. By increasing the polar angle $\theta$ at a fixed azimuthal angle, e.g. $\varphi$= 60°, the state of circular polarization (spin) of the emitted light changes from right-handed to left-handed, shown by red and blue colors respectively. This polar splitting of the modes only occurs when the electron beam traverses the sample at the crossing point of the four sectors (Figure 3e). We attribute this to the fact that the plasmonic crystal is centrosymmetric about point 1, and noncentrosymmetric about point 2.

In general, ascertaining the inversion symmetry of the crystal is attributed to the choice of the origin of the coordinate system. In the structure shown in Figure 2g, depending on the choice of the origin, the crystal could sustain or lack the inversion symmetry (Figure 3c compared to 3e). Thus, positioning the electron beam on sites 1 and 2, the angle-resolved CL maps filtered at the energy of $E$ = 1.55 eV ($\lambda$ ≈ 800 nm) show quite different behaviors. In other words, the superposition of the optical modes excited by the electron beam depend on the electron impact position. As it will be shown in the following section, the spin-orbit interactions break the degeneracy of the plasmonic Bloch modes producing two symmetric and antisymmetric field distributions with different spin vorticities. Thus, site-selective excitations of the sample with the electron beam allow us to decompose the modes. When the electron traverses the sample at the impact position shown in Figure 3c, it excites only the symmetric configuration, while the excitation at the position shown in Figure 3e, could resolve both modes.

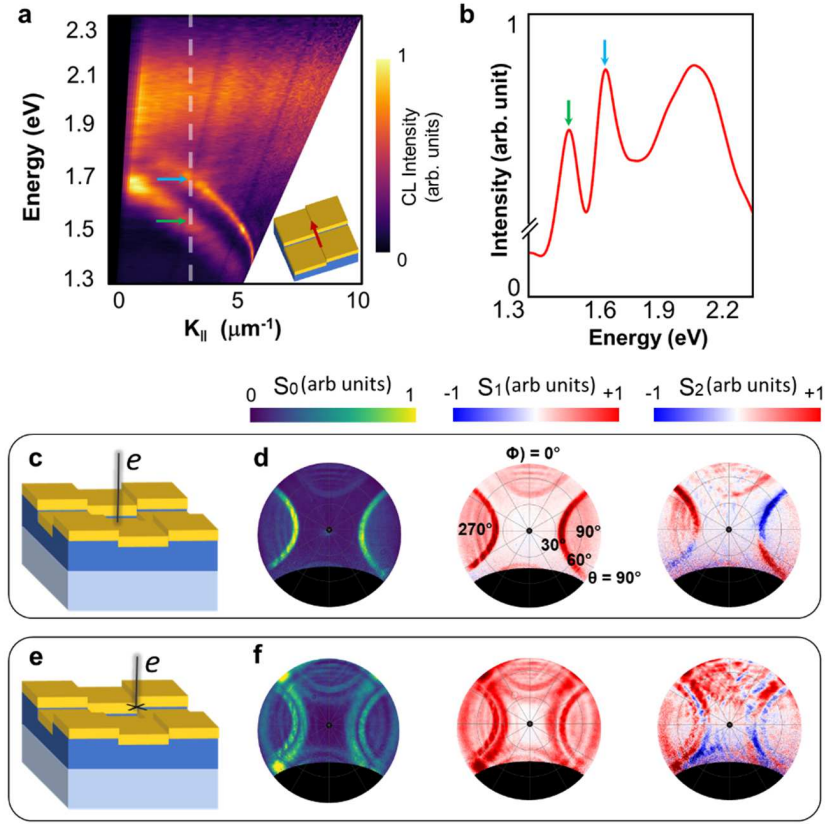

**Figure 3.** Measured cathodoluminescence energy-momentum map, demonstrating the energy splitting of the plasmonic dispersion line. (a) Nondegenerate energy-momentum CL map measured along the direction indicated by a red arrow in the inset. (b) CL spectrum in a constant wavevector indicated by dashed lines in panel (a). The spectral positions of the peaks are indicated by colored arrows. (c, e) Schematic illustration of the sample excited by an electron beam at different positions: (c) at sector 1 of a unit cell, where the ITO thickness is $h_1$ = 65 nm and (e) at the crossing point of the four sectors of a unit cell. (d, f) The Stokes parameters $S_0$, $S_1$ and $S_3$ measured at $E$ = 1.55 eV ($\lambda \approx$ 800 nm) for the electron impact positions indicated in panels (c) and (e), respectively.

Numerical simulations

To better understand the origin of this broken degeneracy and its relation with the inversion symmetry of the structure, the photonic band diagram of the proposed plasmonic crystal is numerically calculated using COMSOL Multiphysics software and compared with the experimentally measured energy-momentum dispersion map. The band diagram for Γ-X path of the first Brillouin zone is depicted in Figure 4a (see Materials and Methods for the details of the numerical simulation), which is in good agreement with the results obtained from the CL measurements (Figure 3a). In the measured energy-momentum map, higher energy modes are manifested as a continuum of optical modes covering a broad range of energies. This is due to the large decay rate of these optical modes associated with the onset of the d-Band transitions in gold. Since the decay rates of the optical modes are not considered in our simulations, the higher energy modes appear as discrete bands in Figure 4a. SOC effects appear in the form of splitting of the energy bands that is clearly observed for the two low-energy modes at Γ point. To compare the optical response of the system at the two different modes, we calculated the z-component of the electric field at points A and B indicated in the bandstructure. Figure 4b exhibits the resulting electric field profiles for A (left panel) and B (right panel) displayed in three different views: (i) *z*-normal plane located below the ITO/Au interface indicated by the dashed lines, (ii) *x*-normal and (iii) *y*-normal planes indicated by the dotted lines in Figure 4b. While the electric field of the first mode penetrates into the substrate, the second mode is highly confined to the ITO/Au interface. Furthermore, the field distribution in the latter case is antisymmetric showing opposite polarities at the two sides of the ITO/Au interface.

We also computed the total angular momentum vorticity of light that is the sum of the spin angular momentum (SAM) and the orbital angular momentum (OAM) vorticities[29,30]:

$$\mathbf{\Omega} = \mathbf{\Omega}_S + \mathbf{\Omega}_O \tag{4}$$

The SAM and OAM vorticities are:

$$\mathbf{\Omega}_S = -\frac{c\varepsilon_0}{8\omega} \text{Im}\left[\nabla^2(\mathbf{E}^* \times \mathbf{E} + \mathbf{H}^* \times \mathbf{H})\right] \tag{5}$$

$$\mathbf{\Omega}_O = \frac{c\varepsilon_0}{4\omega} \text{Im}\left[\nabla \mathbf{E}^* \cdot (\times \nabla)\mathbf{E} + \nabla \mathbf{H}^* \cdot (\times \nabla)\mathbf{H}\right] \tag{6}$$

In equation (6), the scalar product links the field vectors, whereas the vector product relates the gradient operators $\nabla$. Therefore, using the convention specified in ref. (29), we can write:

$$\nabla \mathbf{E}^*(\times \nabla)\mathbf{E} = \nabla E_x^* \times \nabla E_x + \nabla E_y^* \times \nabla E_y + \nabla E_z^* \times \nabla E_z \tag{7}$$

The spin and the total angular momentum vorticities are calculated at the points A and B indicated in Figure 4a. The top views of the SAM vorticities calculated for the first and second modes (Figure 4c) conform well to the field distributions of the corresponding modes, exhibiting rotational flow around the electric field extrema. The spin rotational flow is shown to be right(left)-handed around the maxima (minima) of the electric field. The color map in Figure 4c depicts the size and the direction of the SAM flow, with red and blue denoting right- and left-handed rotations, respectively. The total angular momentum vorticity at point A (Figure 4d, left panel) reveals apparent rotation that is nearly in opposite direction comparing to that calculated at point B (Figure 4d, right panel). The opposite flow of the total angular momentum vorticities obtained for the two split modes conveys the idea that the SOC effects play a significant role in the emergence of the band splitting observed in the noncentrosymmetric plasmonic crystal.

## Discussion

Generally, the bandstructure of symmetric photonic crystals exhibit optical modes that are spin degenerate. To remove the degeneracy of the optical modes and open a bandgap, one has to break either the spatial inversion symmetry or time-reversal symmetry of the crystal. Time-reversal symmetry can be broken by a magnetic field, either intrinsic or external. Alternatively, a gap can be induced when spin-orbit interactions of light occur in an inversion asymmetric crystal. Herein, by combining the experimental results obtained from the cathodoluminescence spectroscopy with those of numerical simulations, we have demonstrated that the spin-orbit coupling effects in a noncentrosymmetric plasmonic crystal can induce splitting of the energy bands giving rise to non-degenerate plasmonic modes with opposite angular momentum distributions. The splitting of the modes was shown to be strongly site-selective, occurring only when the electron beam excites the surface modes at the points wherein the crystal is locally noncentrosymmetric. Interestingly, when we shifted the impact position of the electron to a sector with point symmetry, the splitting vanished due to the local inversion symmetry of the crystal.

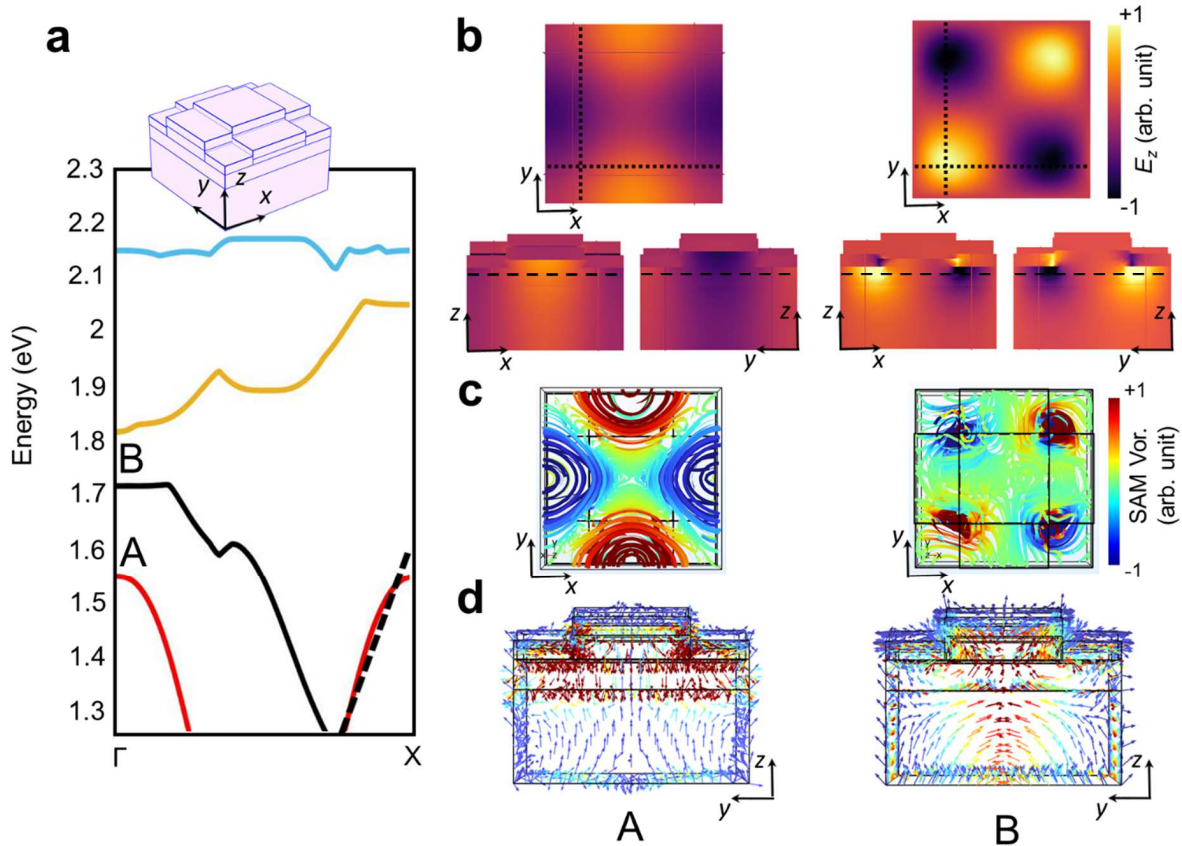

**Figure 4.** Simulated band diagram. (a) The band structures of the noncentrosymmetric plasmonic crystal calculated along Γ-X direction in the reciprocal space. Black dashed line indicates the light line. The insets show unit cells of the corresponding simulated crystals. (b) Top view and cross-sections of the electric field ($E_z$) calculated at points A (left column) and B (right column) shown in the band structure. The dotted and dashed lines indicate the planes in which the cross-sections and top profile are acquired. (c) Spin angular momentum and (d) total angular momentum calculated at points A (left column) and B (right column) shown in the band structure.

Moreover, we determined the polarization state of the light emitted from the sample using angle-resolved CL and dark-field polarimetry. The angular patterns of the Stokes parameters acquired from both methods indicated a polarization-to-direction–coupled emission from the plasmonic crystal. This is attributed to QSHE of light through which the surface plasmon waves propagating in opposite directions carry opposite transverse SAM. Interacting with the periodic crystal, these propagating surface modes scatter to the far-field in different directions based on their polarization (spin) state. This is evident from the angular patterns of the $S_3$ Stokes parameter with a clear spin-dependent direction splitting of the modes. To provide a more precise analysis of the phenomena, we also measured the Stokes parameters for 1D ITO gratings without/with Au layer and compared the results with those obtained from the 2D plasmonic crystal. The fact that a similar spin-dependent direction splitting appears in all samples manifest that the QSHE is not dependent on the symmetry condition of the structure and is intrinsic to all modes with evanescent tails, including waveguiding and surface modes.

Our numerical simulations revealed a band splitting in the *E-k* diagram of the noncentrosymmetric crystal. By calculating the electric field profiles related to the split modes at their $\Gamma$ points, we demonstrated the formation of symmetric and antisymmetric plasmonic modes with different field distributions. These symmetric and antisymmetric modes could be excited selectively by site-selective excitation of the sample with the electron beam. Besides, we conducted numerical simulations to acquire the vorticities of the spin, orbital and total angular momentum in the proposed plasmonic crystal. The total angular momentum vorticities acquired for the two split modes showed rotational behavior with opposite directions. Both SAM and OAM contribute to the total angular momentum distribution leading to the formation of the SOC-induced band splitting, as revealed in the energy-momentum dispersion diagrams obtained from CL measurements and numerical simulations. We deduce that the energy band splitting appears in photonic systems where the inversion symmetry is broken, either locally or globally, while the excitation of the surface plasmon waves induces QSHE of light leading to spin-dependent direction splitting of the light scattered from the sample. In contrast to the QSHE of light that is an inherent property of any waveguiding or surface wave, regardless of the symmetry of the structure, occurrence of band splitting is attributed solely to the symmetry properties of the studied system. Our work may be the first to experimentally demonstrate both effects in noncentrosymmetric plasmonic crystals and can open new pathways for the design of novel spin-orbit plasmonic devices.

## Materials and Methods

**Angle-resolved dark-field polarimetry –** The dark-field polarimetry was carried out using an inverted microscope (Nikon Eclipse Ti2-A). A specialized dark-field condenser enabled diascopic dark-field illumination, i.e. light was only focused onto the sample in an annular ring of large incident angles while the objective was located on the opposite side of the sample. The applied objective was a Nikon S Plan Flour ELWD Series objective with a 60X magnification and a numerical aperture of 0.7, which was smaller than the numerical aperture range of the condenser lens (0.8-0.95). Consequently, only light redirected or emitted by the sample was collected by the condenser. To gain the Stokes parameters, a quarter-wave plate was placed behind the objective followed by a linear polarizer which filtered the collected light by circular or linear polarization, depending on the orientations of their optical axes. Angle-resolved images were obtained by inserting an extra lens, commonly called "Bertrand lens", into the optical path behind the image plane. It focused light rays with the same incident angle in the image plane into one spot of a plane and therefore created a new Fourier image. Optical band pass filters were used to filter the angle-resolved images captured by an Allied Vison Prosilica GC 2450C camera which is an RGB camera optimized for the visible range of light.

**CL spectroscopy and polarimetry –** To perform CL spectroscopy, we used an optical field emission microscope (Zeiss SIGMA) that was equipped with CL compartment made by Delmic B.V. A focused electron beam with an acceleration voltage of 30 keV and a beam current of 11 nA was utilized to excite the surface of the sample. Emitted light was then collected by an off-axis Aluminum parabolic mirror positioned above the sample and was directed to a CCD camera to produce angle-resolved maps (as shown in Figure 1b). The mirror had a focal distance of 0.5 mm, an acceptance angle of 1.46π sr, and dwell time of 35 ms. Angle-resolved patterns formed after the optical beam passed through a color filter to select a specific wavelength (bandwidth of 50 nm). To study the polarization state of the CL emission, we exploited an additional polarizer module including a quarter wave plate (QWP) and a linear polarizer with transmission axes adjusted at angles $\alpha$ and $\beta$, respectively. The angle-resolved CL intensity patterns $I_j$ are measured for six different combinations of ($\alpha$, $\beta$) and the Stokes parameters were calculated by[31]:

$$S_0 = I(0°, 0°) + I(0°, 90°) \tag{8}$$

$$S_1 = I(0°, 0°) - I(0°, 90°)$$

$$S_2 = I(0°, 45°) - I(0°, -45°)$$

$$S_3 = I(45°, 0°) - I(-45°, 0°)$$

where $S_0$ describes the total intensity of the CL emission, $S_1$, $S_2$, and $S_3$ correspond to the linear, diagonal and circular polarization of the light, respectively.

To acquire high quality energy-momentum maps, we used a 220 $\mu$m wide 1D slit in the optical path of the CL emission to select a specific momentum component of the emitted light. Each energy-momentum CL measurement was carried out over a dwell time of 180 s and a diffraction grating was utilized to disperse the light on the camera[28].

**Numerical Simulations** – We conducted numerical simulations using COMSOL Multiphysics software in order to shed light on the optical modes of the plasmonic noncentrosymmetric lattice. The radiofrequency (RF) toolbox of COMSOL was used in a 3D simulation domain, which is based on solving Maxwell's equations in real space and in the frequency-domain using the finite-element method. In order to calculate the band diagrams, a stationary solver was employed to solve a nonlinear eigenvalue problem with the normalized electric field as the eigenvalue. What one obtains with the nonlinear formulation is that the mode normalization performed by the global equation involves setting the domain integral of $\boldsymbol{E}_z \cdot \boldsymbol{E}_z^*$ to unity. The simulation domain was constrained into a single unit cell by the Floquet periodic boundary conditions, and the parametric solver swept the wave vector **k**. The reciprocal lattice vectors of the plasmonic crystal dictate the range of **k**.

### Acknowledgements

This project has received funding from the European Research Council (ERC) under the European Union's Horizon 2020 research and innovation program, Grant Agreement No. 802130 (Kiel, NanoBeam) and Grant Agreement No. 101017720 (EBEAM).

Supporting information for

# Spin-orbit interactions in noncentrosymmetric plasmonic crystals probed by site-selective cathodoluminescence spectroscopy

Masoud Taleb[1], Mohsen Samadi[1], Fatemeh Davoodi[1], Maximilian Black[1], Janek Buhl[2], Hannes Lüder[2], Martina Gerken[2], and Nahid Talebi[1,3*]

[1]Institute of Experimental and Applied Physics, Kiel University, 24098 Kiel, Germany

[2]Integrated Systems and Photonics, Faculty of Engineering, Kiel University, 24143 Kiel, Germany

[3]Kiel, Nano, Surface, and Interface Science, Kiel University, 24098 Kiel, Germany

[*]E-Mail: talebi@physik.uni-kiel.de

## Sample Fabrication

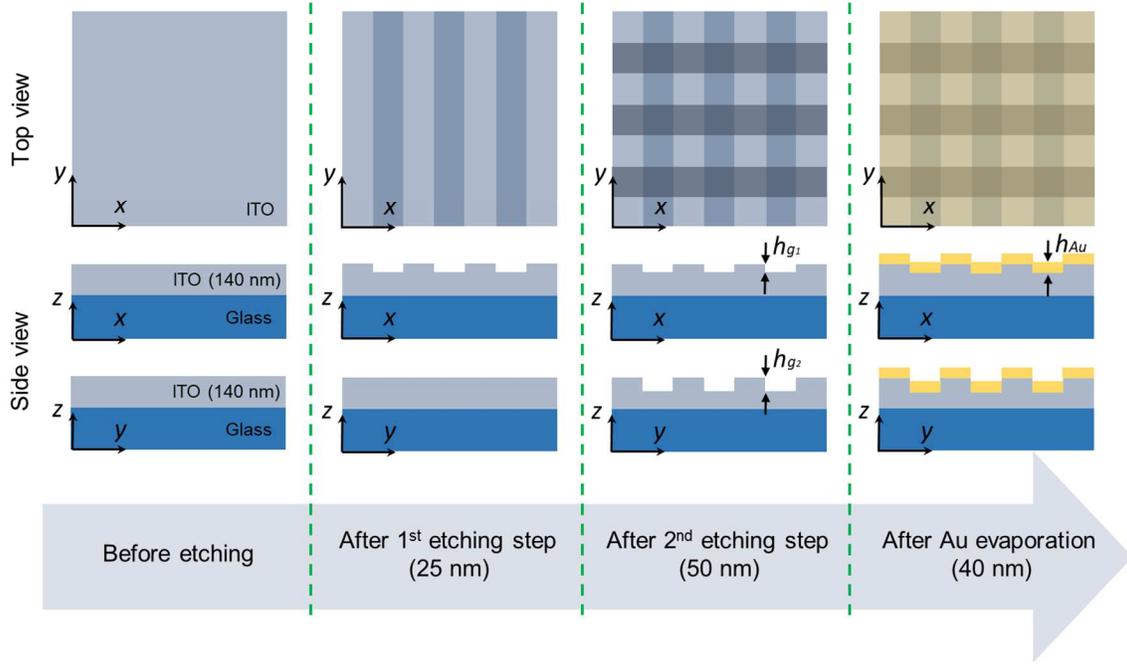

**Figure S5.** Schematic illustration of the fabrication method. A 140 nm thick ITO layer is deposited on a glass substrate. The ITO layer is etched using UV nanoimprint lithography to form a 1D grating along *x*-direction with a groove height of $h_{g1}$ = 25 nm. The sample is then etched along *y*-direction with an etching depth of $h_{g2}$ = 50 nm to form a 2D noncentrosymmetric grating. Finally, the noncentrosymmetric plasmonic crystal is built by evaporating a 40 nm thick layer of Au on the patterned ITO layer[1].

## Field calculation for the evanescent wave

First, we consider an electromagnetic wave propagating along *x*-direction in a medium with permittivity $\varepsilon$ and permeability $\mu$ whose complex electric field can be written as[2,3]:

$$\mathbf{E} = \frac{E_0\sqrt{\mu}}{\sqrt{1+|m|^2}} \begin{pmatrix} 0 \\ 1 \\ m \end{pmatrix} e^{ikx} \tag{S1}$$

where $E_0$ and $k = n\omega/c$ denote the amplitude and the wavenumber of the electromagnetic wave and *n* is the refractive index of the medium. The polarization state of the wave can be expressed as functions of the complex number *m*:

$$\tau = \frac{1-|m|^2}{1+|m|^2}, \quad \chi = \frac{2\mathrm{Re}(m)}{1+|m|^2}, \quad \sigma = \frac{2\mathrm{Im}(m)}{1+|m|^2} \tag{S2}$$

Here, $\tau$, $\chi$, and $\sigma$ are the normalized Stokes parameters exhibiting the degrees of linear, diagonal and circular polarizations respectively. Therefore, *m* = 0 and *m* = ∞ indicate the horizontal and vertical linear polarizations, *m* = ±1 denote the diagonal and anti-diagonal polarizations and *m* = ±i correspond to the right-handed and left-handed circular polarizations.

One can obtain the electric field of an evanescent wave propagating along *x*-direction and decaying along *z*-direction by rotating the Equation (S1) by an imaginary angle $i\varphi$ about *y*-axis. This transformation can be described by the matrix:

$$\widehat{\mathbf{R}}(i\varphi) = \begin{pmatrix} \cosh\varphi & 0 & -i\sinh\varphi \\ 0 & 1 & 0 \\ i\sinh\varphi & 0 & \cosh\varphi \end{pmatrix} \quad (S3)$$

By applying the rotational transformation to the electric field described by the Equation (S1), we have:

$$\mathbf{E} = \frac{E_0\sqrt{\mu}}{\sqrt{1+|m|^2}}\left(-im\frac{\kappa}{k_x}\overline{\mathbf{x}} + \frac{k}{k_x}\overline{\mathbf{y}} + \overline{\mathbf{z}}\right)e^{ik_x x - \kappa z} \quad (S4)$$

where $k_x = k\cosh\varphi$ and $\kappa = k\sinh\varphi$ are the propagation and decay constants respectively. The magnetic field of the evanescent wave can also be calculated using the same method.